\documentclass[twocolums,showpacs,preprintnumbers,amsmath,amssymb]{revtex4}
\usepackage[usenames]{color}
\usepackage{graphicx}
\usepackage{epsfig}
\usepackage{graphicx}% Include figure files
\usepackage{dcolumn}% Align table columns on decimal point
\usepackage{bm}% bold math
\preprint{U. of K. Preprint/2010}
\begin{document}
\baselineskip=22pt
\title{\Large A mass-extended 't Hooft-Nobbenhuis complex transformations and their consequences}
\author{ Arbab I. Arbab\footnote{Email: aiarbab@uofk.edu} and Hisham M. Widatallah\footnote{hisham@ictp.it}}
\affiliation{$^*$Department of Physics,
Faculty of Science, University of Khartoum, P.O. Box 321, Khartoum
11115, Sudan,}
\affiliation{$^\dagger$Department of Physics, College  of Science,
Sultan Qaboos University, Alkhoud 123, Oman}
\date{\today}
\begin{abstract}
We have extended the 't Hooft-Nobbenhuis complex transformations to include  mass.  Under these new transformations, Schrodinger, Dirac, Klein-Gordon and Einstein general relativity equations are invariant. The non invariance of the cosmological constant in Einstein field equations dictates it to vanish thus solving the longstanding cosmological constant problem.
\end{abstract}
\maketitle
\section{Introduction}
Recently  't Hooft and Nobbenhuis have introduced  complex space-time transformations under which the invariance of the ground - state  would associate the vacuum state to a zero cosmological constant. These transformations are such that the space-time coordinates $x^\mu\rightarrow ix^\mu$. Under these transformations, the Hamiltonian of a nonrelativistic particle $H\rightarrow -H$ and the boundary conditions of the physical states do not become invariant. This is because while the real part of the states goes to zero as $x\rightarrow \infty$, the imaginary part doesn't. Consequently, all states except the ground state ($\psi (x)=\rm const.$) will break this symmetry. In the context of quantum theory, hermiticity, normalization and boundary conditions will not transform as in the usual symmetry transformations for a non-relativistic particle.

In as  much as in quantum mechanics energy and momentum are described by $p=-i\hbar \vec{\nabla}$ and $E=i\hbar\frac{\partial}{\partial t}$,  any coordinates transformations will inevitably transform $p$ and $E$. Moreover, in relativity theory mass and energy are related. Hence, the coordinate transformations will necessarily require the mass transformations too. This latter transformations have been overlooked by 't Hooft and Nobbenhuis in their complex transformations.

We argue that the  inclusion of  mass transformation will resolve the shortcomings of the theory pertaining to the hermiticity, normalization and boundary conditions.  The  resulting complex transformations,  referred to   the 't Hooft-Nobbenhuis mass-extended transformations, lead to interesting physics when applied to Schrodinger, Dirac, Einstein general relativity and Maxwell equations. Under these transformations, the hermitian operators are transformed into antihermitian operators. Moreover, the wavefunctions that are periodic in real spaces are also periodic in imaginary space. These pretty transformations urge us to formulate our laws of nature in a complex space instead of the present real space.

\section{The mass-extended 't Hooft-Nobbenhuis transformations}
't Hooft and Nobbenhuis have recently introduced a complex space-time transformations, and identified it as a symmetry of laws of nature.
In quantum mechanics, the space-time transformations will transform momentum ($p$) and energy ($E$) since the latter are expressed by
\begin{equation}
\vec{p}=-i\hbar\vec{\nabla}\,,\qquad\qquad E=i\hbar\frac{\partial}{\partial t}\,.
\end{equation}
From the theory of relativity, one knows that mass ($m$) and energy  are related by
\begin{equation}
E=mc^2\,,\qquad\qquad p=mv\,.
\end{equation}
Hence, Eq.(1) and (2) will yield
\begin{equation}
E'= -iE\,,\qquad \vec{p}\,'=-i\vec{p}\,,\qquad m'= -im\,.
\end{equation}
Therefore, the full complex transformations will become
\begin{equation}
\vec{r}\,'= -i\vec{r}\,,\qquad t'= -it\,,\qquad m'= -im\,.
\end{equation}
We refer to these transformations as 't Hooft-Nobbenhis
mass-extended transformations. These transformations  do not alter
the Einstein mass-energy equation, $E=\sqrt{c^2p^2+m_0^2c^4}$ and
the Lorentz transformations. According to Eq.(4), one notices that
\begin{equation}
E't'= Et\,,\qquad \vec{p}\,'\cdot \vec{r}\,'=  \vec{p}\cdot
\vec{r}\,,\qquad \hbar\,'=\hbar\,,\qquad c\,'=c
\end{equation}
so that the wave phase does not change. This guarantees the normalization and boundary conditions of all physical states. Hence, Eq.(4) is a symmetry of laws of nature.
\section{Schrodinger equation}
Applying the transformations in Eq.(4) to Schrodinger equation
\begin{equation}
i\hbar\frac{\partial \psi}{\partial t}=-\frac {\hbar^2}{2m}\nabla^2\psi+V\psi
\end{equation}
shows that Schrodinger equation is invariant provided $V(ix)=-iV(x)$. The probability and current densities in Schrodinger formalism are defined by
\begin{equation}
\rho=\psi^*\psi\,,\qquad\vec{J}=\frac{\hbar}{2mi}\left(\psi^*\vec{\nabla} \psi-\psi\vec{\nabla} \psi^*\right)
\end{equation}
so that the continuity equation reads
\begin{equation}
\vec{\nabla}\cdot\vec{J}+\frac{\partial \rho}{\partial t}=0\,.
\end{equation}
Applying the transformations (4) in Eq.(7) yields
\begin{equation}
\vec{J}\,'=i\vec{J}\,,\qquad\qquad \rho\,'=i\rho\,.
\end{equation}
This implies that the continuity equation, Eq.(8), is invariant under the transformation in Eq.(4) too.
\section{Non relativistic particle motion}
Following 't Hooft and Nobbenhuis , we discuss here a one dimensional motion of a non-relativistic particle. Consider the Hamiltonian
\begin{equation}
H=\frac{p^2}{2m}+V(x)\,.
\end{equation}
Under the transformations in Eq.(4), the above equation yields
\begin{equation}
H'=\frac{-p^2}{-2mi}+V(ix)=-i\left(\frac{p^2}{2m}+V(x)\right)=-iH
\end{equation}
where, $V(ix)=-iV(x)$. This can be realized for certain potentials.

If we now consider a harmonic oscillator where the Hamiltonian is given by
\begin{equation}
H=\hbar\,\omega\,(a^+a+\frac{1}{2})
\end{equation}
where\footnote{$[\omega]=T\,^{-1}.$}
\begin{equation}
a=\sqrt{\frac{m\omega}{2\hbar}}\left(x+\frac{ip}{m\omega}\right)\,,\qquad a^+=\sqrt{\frac{m\omega}{2\hbar}}\left(x-\frac{ip}{m\omega}\right).
\end{equation}
Under the transformations in Eq.(4), one has
\begin{equation}
a'= -a\,,\qquad a'^+=- a^+\,,\qquad H'= -iH\,,\qquad m'\omega'=-m\omega\,.
\end{equation}
The state wavefunction, $\psi_n(x)$, which is given by
\begin{equation}
\psi_n(x)=\frac{1}{\sqrt{2^nn!}}\left(\frac{m\omega}{\pi\hbar}\right)^{1/4} \exp(-\frac{m\omega}{2\hbar} x^2)\,H_n(\sqrt{\frac{m\omega}{\hbar}}\,x)
\end{equation}
where
\begin{equation}
H_n(x)=(-1)^ne^{x^2}\frac{d^n}{dx^n}\left(e^{-x^2}\right)\,,\qquad n=0, 1, 2, \ldots
\end{equation}
transforms as, $\psi\,'_n(x)= (-1)^n \sqrt{i}\,\psi_n(x)$. This is also consistent with Eqs.(7) and (9). Hence, the boundary  and  normalization conditions are satisfied for all states. This is unlike the original 't Hooft-Nobbenhuis
transformations, where these two properties of the wavefunction are lost. Moreover, the transformed energy and momentum operators are
antihermitian.
\section{Classical scalar field}
Consider a real  scalar field $\Phi(x)$ defined by the  Lagrangian density
\begin{equation}
{\cal L}=-\frac{1}{2}(\partial_\mu\Phi)^2-V(\Phi)\,,\qquad V(\Phi)=\frac{1}{2}m^2\Phi^2+\lambda \Phi^4
\end{equation}
and the Hamiltonian density
\begin{equation}
{\cal H}=\frac{1}{2}\,\Pi^2+\frac{1}{2}(\nabla\Phi)^2+V(\Phi)\,,\qquad \Pi(x)=\partial_0\Phi
\end{equation}
where $\lambda$ is a constant. Now if $\Phi(x)$ transforms as
\begin{equation}
\Phi'(x)\equiv\Phi(ix)=-i\Phi(x)\qquad\Rightarrow\qquad \Pi'(ix)=-\Pi(x)
\end{equation}
the above Lagrangian and Hamiltonian will be invariant, i.e., $\cal L'=\cal L$ and $\cal H'=H$. Moreover, the action $S=\int {\cal L}\, d^4x=S\,'$.

\section{Maxwell equations}
Defining the electromagnetic tensor $F_{\mu\nu}$ as [3]
\begin{equation}
F_{\mu\nu}=\partial_\mu A_\nu-\partial_\nu A_\mu
\end{equation}
Maxwell equations read
\begin{equation}
\partial_\mu F^{\mu\nu}=\mu_0 J^\nu\,,\qquad \partial_\mu F_{\nu\lambda}+\partial_\nu F_{\lambda\mu}+\partial_\lambda F_{\mu\nu}=0\,.
\end{equation}
Under the transformations in Eq.(4), the charge ($q$), current ($I$) and vector potential $A_\mu$ transform as
\begin{equation}
q'= q\,,\qquad I'=-iI\,,\qquad A'_{\mu}= -iA_\mu\,.
\end{equation}
 Therefore, the electromagnetic tensor, the charge and current densities transform as
\begin{equation}
F\,'_{\mu\nu}=-F_{\mu\nu}\,,\qquad \vec{J}\,'=i\vec{J}\,,\qquad \rho'= i \rho\,.
\end{equation}
Hence, Maxwell equations are invariant under Eq.(4). If we extend
our analysis to Yang-Mills field, the Lagrangian will involve
quadratic, cubic and quartic couplings of the filed $A_\mu$. In
this case, one has $F_{\mu\nu}\rightarrow F^a_{\mu\nu}$, where
\begin{equation}
F^a_{\mu\nu}=\partial_\mu A^a_\nu-\partial_\nu A^a_\mu+if^{abc}A^b_\mu A^c_\nu
\end{equation}
where $f^{abc}$ are the structure constants, we will get an
invariant Lagrangian, since
\begin{equation}
F^a_{\mu\nu}\,'=-F^a_{\mu\nu}\,.
\end{equation}
\section{Quantum electrodynamics-QED}
The QED Lagrangian density for a free particle with rest mass
$m_0$ is given by [2]
\begin{equation}
{\cal L}=i\hbar \,\bar{\psi}\,\gamma^\mu D_\mu\psi-m_0c\,\bar{\psi}\,\psi-\frac{1}{4}F_{\mu\nu}F^{\mu\nu}
\end{equation}
where $D_\mu=\partial_\mu-i\frac{e}{\hbar}A_\mu\,.$ This
Lagrangian is invariant under Eq.(4) and (17), viz., $\cal L'=\cal
L$, provided that $\bar{\psi}\,'\,\psi'=i\,\bar{\psi}\,\psi$,
where we have assumed $\gamma\,'^\mu=\gamma^\mu$. This can be
satisfied if $\psi\,'=\sqrt{i}\,\psi$ and $\bar{\psi}\,'=\sqrt{i}\,\bar{\psi}$. This ushers in the direction that the probability and current density transform as $ \rho\,'=i\,\rho$ and $\vec{J}\,'=i\,\vec{J}$. This is in agreement with the transformations in Eqs. (9) and (17).
\section{General theory of relativity}
Einstein equations for general relativity are [4]
\begin{equation}
R_{\mu\nu}-\frac{1}{2}R\,g_{\mu\nu}=8\pi\,G T_{\mu\nu}\,.
\end{equation}
The particle equation of motion is given by the geodesic equation
\begin{equation}
\frac{d^2x^\mu}{d\tau^2}+\Gamma^\mu_{\nu\lambda}u^\nu
u^\lambda=0
\end{equation}
where $\Gamma^\mu_{\nu\lambda}$ are the Christoffel symbols, $\tau$ and $u^\mu$ are the proper time and velocity, respectively.
Under the transformations (4), Eq.(28) yields
\begin{equation}
\Gamma^\mu_{\nu\lambda}\,'=-i\,\Gamma\,^\mu_{\nu\lambda}\,.
\end{equation}
Using the transformations in Eq.(4), one finds that\footnote{$[G]=M^{-1}L^3T^{-2}$ and [$\rho_m]=ML^{-3}$.}
\begin{equation}
\rho'_m=\rho_m\,,\qquad G'=-G\,,\,\qquad R'_{\mu\nu}=-R_{\mu\nu}\,,\qquad T'_{\mu\nu}=T_{\mu\nu}
\end{equation}
where $\rho_m$ is the matter density and $G$ is Newton constant. Hence, Einstein general relativity equations are invariant under the transformations in Eq.(4). However, the existence of the cosmological in the Einstein field equations will violate the invariance. The way out of this is that the cosmological constant must be zero.  Thus, the vanishing of the cosmological constant is thus because its existence violates the symmetry defined in Eq.(4). But if the cosmological constant has to be present in the Einstein field equations, it  must change sign under the transformations in Eq.(4). In this case, the Einstein field equations are invariant. This global invariance is a quite interesting merit that  the original 't Hooft- Nobbenhuis transformations had wished for. Hence, the reason for the vanishing of the cosmological constant is now understood.
\section{Concluding remarks}
We have extended in this work the complex space-time transformation postulated by 't Hooft and Nobbenhuis to include mass. This extended transformation remedied the problems of the original transformations. These transformations can be considered as special case of scale transformation. We have found n in this work that all physical laws are invariant under the complex space-time and mass transformations. Hence, the extended complex transformations are the symmetry of laws of nature.
\section*{References}
\hspace{-0.55cm}  Gerard 't Hooft, Stefan Nobbenhuis,   \emph{Class. Quantum Grav}. 23, 3819 (2006).\\
 Bjorken, J. D., and Drell, S. D., \emph{Relativistic Quantum Mechanics}, McGraw-Hill (1964).\\
 Jackson,  D., \emph{Classical Electrodynamics}, John Wiley \& Sons Inc. (1962).\\
 Weinber, S., \emph{Gravitation and cosmology}, John Wiley \& Sons Inc. (1972).\\
\end{document}